\documentclass[a4paper,11pt]{article}
\usepackage{pos}
\usepackage{smartdiagram,multirow,hhline}

\newcommand{\dd}{\mathrm{d}}
\newcommand{\DD}{\mathrm{D}}

\newcommand{\w}{\wedge}

\newcommand{\be}{\begin{equation}}
\newcommand{\ee}{\end{equation}}
\newcommand{\sfrac}[2]{{\textstyle\frac{#1}{#2}}}
\def\nn{\nonumber}
\def \bea{\begin{eqnarray}} 
\def\eea{\end{eqnarray}}

\def\bi{\begin{itemize}} 
\def\ei{\end{itemize}}

\makeatother

\def\a{\alpha} \def\b{\beta}  \def\G{\Gamma} \def\d{\delta} \def\D{\Delta}
\def\e{\epsilon} 
   
\def\l{\lambda}  
    
\def\s{\sigma} \def\S{\Sigma}  \def\th{\theta}

\def\one{\mbox{1 \kern-.59em {\rm l}}}
\smartdiagramset{bubble center node size=6cm, bubble node size=2cm, planet size=0.5cm, distance center/other bubbles=0.6cm}
\title{Twisted R-Poisson Sigma Models}
 \ShortTitle{Twisted R-Poisson Sigma Models}

\author*[a]{Athanasios Chatzistavrakidis}

\affiliation[a]{Division of Theoretical Physics, Rudjer Bo\v skovi\'c Institute,\\
   Bijeni\v cka 54, 10000 Zagreb, Croatia}

\emailAdd{athanasios.chatzistavrakidis@irb.hr}

\abstract{The AKSZ construction was developed as a geometrical formalism to find the solution to the classical master equation in the BV quantization of topological branes based on the concept of QP manifolds. However, the formalism does not apply in presence of Wess-Zumino terms, as demonstrated recently by Ikeda and Strobl in the simplest example of WZW-Poisson sigma models. In this contribution, we review a class of topological field theories in arbitrary dimensions, the twisted R-Poisson sigma models, which suitably generalize Poisson or twisted Poisson sigma models. Their relation to differential graded manifolds and higher geometry is discussed and we sketch how to identify the solution to the classical master equation even though the target space does not have a QP structure.}

\FullConference{%
  Corfu Summer Institute 2021 "School and Workshops on Elementary Particle Physics and Gravity"\\
  29 August - 9 October 2021\\
  Corfu, Greece
}

\begin{document}
\maketitle

\section{Introduction and motivation}

Twisted and bi-twisted R-Poisson sigma models \cite{Chatzistavrakidis:2021nom} are topological field theories in diverse dimensions that constitute a natural extension of the two-dimensional Poisson \cite{SchallerStrobl,Ikeda} and 3-form-twisted (or Wess-Zumino-Witten) Poisson sigma models \cite{Klimcik:2001vg}. Their underlying geometrical structure encompasses and extends Poisson geometry to the so-called R-Poisson geometry and its twisted versions.{\footnote{Further details on the geometry and the extension to Dirac structures may be found in Ref. \cite{Ikeda:2021rir}.}} R-Poisson geometry comprises a manifold $M$ equipped with a Poisson structure $\Pi$ (an antisymmetric bivector) and an additional $R$-structure (an antisymmetric multivector of order $p+1$), such that the antisymmetric multivector of order $p+2$ obtained from the natural Schouten-Nijenhuis bracket of multivector fields $[\Pi,R]_{\text{SN}}$ vanishes. If $M$ is additionally endowed with a closed $p+2$ form $H$, then this Schouten-Nijenhuis bracket can be nonvanishing, instead being controlled by the contraction of $H$ with the Poisson bivector $p+2$ times. This $(M,\Pi,R,H)$ geometry is called twisted R-Poisson. At the field theory level of twisted R-Poisson sigma models, it governs the invariance of the sigma model action functional under gauge transformations and the on-shell closure of the gauge algebra. More generally, it is the main consistency condition for the classical master equation to hold, when one performs the Batalin-Vilkovisky (BV) quantization of the topological field theory. 

Why, however, should we be interested in structures such as the one described in the previous paragraph? The main motivation to introduce twisted R-Poisson sigma models was the work of Ikeda and Strobl \cite{Ikeda:2019czt}, who demonstrated that the AKSZ construction of topological field theories does not work for the WZW-Poisson sigma model in two dimensions and moreover they found the solution to the classical master equation beyond AKSZ. The AKSZ construction \cite{Alexandrov:1995kv} was developed as a general geometrical method to find the solution to the classical master equation in the BV quantization of topological strings and higher branes. Its backbone is a QP structure, a differential graded (namely, one equipped with a cohomological vector field $Q$) symplectic supermanifold, which once identified it can serve as the target space in a sigma model with a differential graded or Q-manifold as source spacetime. 

Prototypically, one considers a Poisson manifold $(M,\Pi)$ and its cotangent bundle $T^{\ast}M$ with a Lie algebra structure on its sections given by the Koszul-Schouten bracket of differential 1-forms. Anchoring this structure to the tangent bundle $TM$ by means of a smooth map $\Pi^{\sharp}:T^{\ast}M\to TM$ induced by the bivector $\Pi$, assigns a so-called Lie algebroid structure to the cotangent bundle. Furthermore, using Vaintrob's isomorphism between Lie algebroids on a vector bundle $E$ and Q-manifolds on the degree shifted vector bundle $E[1]$ \cite{Vaintrob}, where the fiber coordinates are assigned degree 1 as appears in the brackets,  one considers the Q-manifold $T^{\ast}[1]M$ with its accompanying cohomological vector field of degree 1 and its natural odd symplectic P-structure as a cotangent bundle. Realizing that the symplectic form of the P-structure is invariant along the flow of the cohomological vector field Q, there exists a QP structure on $T^{\ast}[1]M$ and the AKSZ construction may be applied. The result is the BV action of the Poisson sigma model, i.e. the solution to its classical master equation, and of the topological A-model after gauge fixing. 

However, in the presence of a 3-form $H$ on $M$, although both Q and P structures continue to co-exist on $T^{\ast}[1]M$, nevertheless the strict compatibility of the two is lost. Indeed the change in the symplectic form along the flow of the cohomological vector field ceases to be zero and it becomes an exact 2-form controlled by the contraction of $H$ with the $H$-twisted Poisson bivector $\Pi$ two times. Thus the QP structure is obstructed and this is the basic reason that the AKSZ construction does not apply---and even a naive generalization of it does not yield a correct result, as discussed in \cite{Ikeda:2019czt}.  

Clearly, the reason for the above conundrum is the presence of a Wess-Zumino term in the two-dimensional WZW-Poisson sigma model. One may then ask how special is this situation. In other words, are there any other examples of topological field theories where the AKSZ construction does not apply and nevertheless the classical master equation can be solved and be given a geometric interpretation? Twisted R-Poisson sigma models are a large class of possible such examples. For instance, the AKSZ construction in three dimensions leads to an extension of Chern-Simons theory known as the Courant sigma model \cite{Ikeda:2000yq,Ikeda:2002wh,Hofman:2002jz,Roytenberg:2006qz}. 
Suppose that we consider this theory on a membrane without boundary and turn on a Wess-Zumino term, which in the present case it would be a 4-form supported on a three-brane with boundary this membrane. Such 4-form-twisted Courant sigma models were considered before in \cite{Hansen:2009zd}, but their BV action has not been studied. In our approach, this is fully addressed and the BV action can be found, at least in the twisted R-Poisson subclass of twisted Courant sigma models \cite{CIS}. 

Recalling that the BV formalism was developed to quantize gauge systems with so-called open gauge algebras, i.e. gauge algebras that close only on-shell, and reducible systems where not all gauge parameters are independent, it is useful to mention that twisted R-Poisson sigma models are examples of such theories. They are multiple stages reducible, since they contain higher degree differential forms in their field content and moreover the algebra of gauge transformations contains terms proportional to the classical equations of motion. One unorthodox feature that is usually absent in discussions of the BV formalism is that the gauge algebra is in fact proportional to \emph{products} of equations of motion, i.e. it is nonlinear in them. Apart from being rather unusual, there is nothing wrong with this property, since the gauge algebra is still on-shell closed. However, its treatment becomes more complicated. 

From the perspective of differential graded manifolds, twisted R-Poisson geometry is realized on the higher degree-shifted cotangent bundle ${\cal M}=T^{\ast}[p]T^{\ast}[1]M$. This is a Q-manifold with four types of coordinates of degrees $0, 1, p-1$ and $p$ respectively. Indeed, suppose that the ordinary manifold $M$ is equipped with (ordinary, bosonic, degree-$0$) coordinates $x^{i}$. Then a coordinate system on the degree-shifted cotangent bundle ${\cal M}_0=T^{\ast}[1]M$ contains the coordinates $x^{i}$ together with anticommuting, degree-$1$, fermionic coordinates $a_i$. From another point of view, thinking of the cotangent bundle as a phase space, $a_i$ are the graded momenta of the coordinates $x^{i}$. In a second step, one considers the higher phase space ${\cal M}=T^{\ast}[p]{\cal M}_0$. A coordinate system on it consists of the coordinates $x^{i}, a_i$ on the base ${\cal M}_0$ along with the corresponding graded momenta such that the symplectic form is of degree $p$. These are then the degree-$p$ momenta $z_i$ of $x^{i}$ and the degree-$(p-1)$ momenta $y^{i}$ of $a_i$. As we will discuss below, this perspective is particularly useful to understand the obstruction to QP-ness discussed above. 

 \begin{figure}
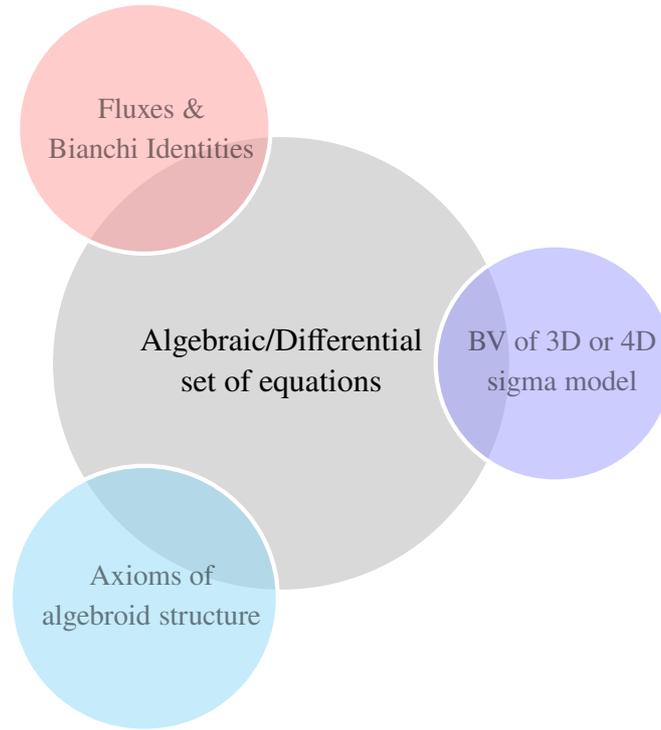
 \begin{center}	\smartdiagram[bubble diagram]{Algebraic/Differential \\ set of equations,
			Fluxes \&  \\ Bianchi Identities , Axioms of \\ algebroid structure, BV of 3D or 4D \\ sigma model} 
			\end{center} \caption{Triple points}\label{t1} \end{figure}

A further motivation to study twisted R-Poisson sigma models regards world volume perspectives to string theory backgrounds with general fluxes. Notably, it has been shown that in a number of cases of interest, there exists ``triple points'' (see \cite{Chatzistavrakidis:2018ztm} and references therein for the gradual development of this perspective),  that is sets of algebraic and differential equations that may be interpreted in three distinct ways: (i) as the equations governing the gauge structure and the classical master equation of certain world volume topological quantum field theories (topological brane sigma models), (ii) as consistency conditions (Bianchi identities) in flux compactifications of certain target space theories (10 or 11-dimensional supergravities or double field theory), and (iii)  as the local coordinate form of the axioms of certain higher structures such as Courant or DFT algebroids. This may be sketched as in Figure \ref{t1}. 

Apart from being motivated by this interrelation, the work reviewed in this contribution also clarifies a further issue, which will be discussed briefly here---details are found in the original reference \cite{Chatzistavrakidis:2021nom}. In a number of works relating membrane and threebrane sigma models to string and M-theory backgrounds \cite{Mylonas:2012pg,Chatzistavrakidis:2019seu,Chatzistavrakidis:2018ztm,Bessho:2015tkk,Heller:2016abk}, the action functionals, gauge or BRST transformations and field equations are defined on a local patch only. Following the reasoning on target space covariant formulations discussed in \cite{Ikeda:2019czt} for the WZW-Poisson sigma model, one can now express such topological sigma models in a fully covariant form and only in terms of tensorial quantities that are associated to the notion of ``$E$-geometry'', i.e. the study of $E$-connections together with their $E$-torsion and $E$-curvature.   
The latter constitutes the application of ordinary differential-geometric concepts to algebroids $E$ (in the present context, Lie algebroids in particular.)

The rest of this contribution is structured as follows. Section \ref{sec2} contains a brief discussion of some essential features of the WZW-Poisson sigma model, in particular its description on a local patch and its covariant formulation in terms of $E$-connections with $E=T^{\ast}M$. The same features are then discussed for the twisted R-Poisson sigma models that are introduced in the same section. We complement this with a discussion about further deformations of twisted R-Poisson sigma models in special dimensions and a clarification of their relation to Poisson and twisted Poisson sigma models in two dimensions. Section \ref{sec3} contains a discussion on the differential graded manifold perspective to twisted R-Poisson structures and clarifies the lack of QP structure in the presence of Wess-Zumino terms. Building on this, we then recall the main steps of the AKSZ construction in the examples of the 2D Poisson and 3D Courant sigma model, as well as the main steps required to find the solution to the classical master equation beyond AKSZ through the traditional BV approach. We conclude in section \ref{sec4} with a number of comments, take-home messages and an outlook.

\section{Twisted R-Poisson sigma models \& beyond} 
\label{sec2} 

\subsection{Twisted Poisson in 2D}
\label{sec21}

A Poisson structure can be described in several equivalent ways. For our purposes it is useful to think of it as the pair $(M,\Pi)$ of a  smooth manifold $M$ and an antisymmetric 2-vector $\Pi\in \Gamma(\wedge^2TM)$, the Poisson structure. That $\Pi$ is a Poisson structure is expressed as the vanishing of the Schouten-Nijenhuis bracket of $\Pi$ with itself, 
\be \label{Poisson}
[\Pi,\Pi]_{\text{SN}}=0\,.
\ee 
We recall that the Schouten-Nijenhuis bracket is the natural extension of the Lie bracket of vector fields to multivector fields. For $v$'s and $u$'s being vector fields with Lie bracket $[v_i,u_j]$, it takes a $p$-vector and a $q$-vector and gives a $(p+q-1)$-vector as 
\be 
[v_1\dots v_p,u_1\dots u_q]_{\text{SN}}=\sum_{i,j}(-1)^{i+j}[v_i,u_j]v_1\dots v_{i-1} v_{i+1}\dots v_{p} u_1\dots  u_{j-1}u_{j+1}\dots u_q\,,
\ee 
with wedge products understood on the right hand side. For example, 
Eq. \eqref{Poisson} reads 
\be 
\Pi^{[i|l|}\partial_l\Pi^{jk]}=0\,,\ee
which is nothing but the Jacobi identity for the Poisson bracket of functions $f$ and $g$, 
\be 
\{f,g\}=\Pi(\dd f,\dd g)\,.
\ee 

A Poisson structure can be twisted by a 3-form. A 3-form twisted Poisson manifold is the triple $(M,\Pi,H_3)$ of a smooth manifold equipped with an antisymmetric 2-vector $\Pi$ and a 3-form $H_3$ such that \cite{Severa:2001qm}
\bea \label{twistedPoisson}
\frac 12 [\Pi,\Pi]_{\text{SN}}=\langle\otimes^3\Pi,H_3\rangle \quad \text{and} \quad \dd H_3=0\,.
\eea
The failure of $\Pi$ to be Poisson is controlled by the closed 3-form $H_3$ and measured by its triple contraction with $\Pi$. 

Both Poisson and twisted Poisson structures bear a relation to the structure of a Lie algebroid $(M,\rho,[\cdot,\cdot]_E)$. This is a vector bundle $E$ over a smooth manifold $M$, anchored to the tangent bundle via a smooth bundle map $\rho:E\to TM$ and equipped with a Lie algebra structure on its sections. In other words, there exists a binary, skew-symmetric operation, a Lie bracket $[e,e']_{E}$ for sections $e,e'\in\G(E)$, which satisfies the Leibniz rule 
\be 
[e,fe']_E=f[e,e']_E+\rho(e)f\, e'\,.
\ee 
Poisson structures and their twisted extension are related to this concept when one makes the choice that the vector bundle is the cotangent bundle, namely $E=T^{\ast}M$, the map $\rho$ is the ``musical'' isomorphism $\Pi^{\sharp}:T^{\ast}M\to TM$ induced by a (possibly twisted, in the above sense) Poisson structure on $M$ and the bracket on sections of the cotangent bundle (that is, 1-forms) is the (twisted) Koszul bracket 
\be 
[e,e']_K={\cal L}_{\Pi^{\sharp}(e)}e'-{\cal L}_{\Pi^{\sharp}(e')}e-\dd (\Pi(e,e'))+H_3(\Pi^{\sharp}(e),\Pi^{\sharp}(e'))\,.
\ee 
Then $(T^{\ast}M,\Pi^{\sharp},[\cdot,\cdot]_{K})$ is a Lie algebroid if and only if the defining conditions \eqref{twistedPoisson} hold. 

A topological sigma model in two dimensions can be constructed on the basis of twisted Poisson geometry. It comprises a field content of real scalar fields $X^{i}=X^{i}(\s)$ accompanied with worldsheet 1-forms $A_i=A_{i\a}(\s)\dd \s^{\a}$, where $\s^{\a}, \a=0,1$ are local coordinates on the worldsheet $\S_2$. In more geometric terms, the fields are components of 
\be 
X\in C^{\infty}(\S_2,M) \quad \text{and} \quad A\in \Omega^{1}(\S_2,X^{\ast}T^{\ast}M)\,.
\ee 
We observe that the 1-forms $A_i$ take values in the pull-back bundle of the cotangent bundle of $M$ through the pull-back map of the sigma model map $X:\S_2\to M$.
The action functional of the topological field theory is \cite{Klimcik:2001vg}
\be \label{SHPSM}
S_{\text{HPSM}}=\int_{\S_{2}}\left(A_i\w\dd X^{i}+\frac 12 \, \Pi^{ij}(X)\,A_i\w A_j\right)+\int_{\S_3}X^{\ast}H_3\,. 
\ee  
The twist $H_3$ of the geometrical structure manifests itself as a Wess-Zumino term in the theory, supported on an open membrane with boundary the worldsheet $\S_2$ and obeying certain conditions that render the theory well-defined as a 2D one, see e.g. \cite{Figueroa-OFarrill:2005vws}. 

The 3-form-twisted Poisson sigma model described above is an example of a gauge theory with gauge algebra that does not close off-shell, in other words of one where the commutator of two gauge transformations result in a trivial gauge transformation. The gauge symmetries of the theory are 
\bea  
\d_\e X^{i}=\Pi^{ji}\e_j \quad \text{and} \quad \d_{\e}A_i=\dd\e_i+\partial_i\Pi^{jk}A_j\epsilon_k+\frac 12\, \Pi^{jk}H_{ijl}(\dd X^{l}-\Pi^{lm}A_m)\epsilon_k\,.
\eea  
The quantities $\partial_i\Pi^{jk}$ that appear in the gauge transformation of the 1-forms $A_i$ are essentially the structure functions of the gauge algebra as well as the ones of the Lie algebroid (Koszul) bracket in a coordinate system---we thus see that apart from having a so-called ``open'' gauge algebra, the theory at hand has a so-called ``soft'' gauge algebra too, i.e. one where the structure ``constants'' are not constant. The gauge transformation of $A_i$ is then just the ordinary one for nonlinear gauge theory, save the ultimate $H_3$-dependent term. This last term appears due to the Wess-Zumino term of the theory and it cancels its contribution to $\d_{\e}S_{\text{HPSM}}$.  We will see below that the specific combination of the 1-forms $\dd X^{i}$ and $A_i$ appearing in it, are in fact a special case of a general combination for higher-dimensional topological sigma models with an underlying Poisson or twisted Poisson structure.

The classical field equations of the theory read 
\be 
F^{i}:=\dd X^{i}+\Pi^{ij}A_j=0\,,
\qquad G_i:=\dd A_i+\frac 12 \, \partial_i\Pi^{jk}A_j\w A_k+\frac 12 \, H_{ijk}\dd X^{j}\w\dd X^{k}=0\,.
\ee 
We observe that the first one appears in the gauge transformation of $A_i$. However, one should note that it is not the parenthesis of the second term. Therefore, not all $H$-dependence of $\d_{\e}A_i$ resides in a field equation dependent term. 

Before moving on, it is worth discussing the invariant geometrical meaning of what was presented so far, following the reasoning in \cite{Ikeda:2019czt}. In a discussion like this, one should consider the action functional itself, the gauge transformations and the field equations. However, in the present case the action functional is essentially already taken case of. It suffices to notice that $A\w\dd X$ and $\Pi(A\w A)$ are well defined 2-forms, see for example \cite{Fulp:2002fm}. In fact, the same holds for the gauge transformation on $X$ and the field equation of $A$, which may be written as 
\be 
\d_{\e}X=\Pi(\e,\cdot) \quad \text{and} \quad F=\dd X+\Pi(\cdot,A)\,,
\ee
and we have indicated the slot where the contraction takes place. 

What remain are the gauge transformation of $A$ and the field equation for $X$. Observing that both of them contain a partial derivative on the components of the tensor $\Pi$, it is clear that we need a connection to turn this object into a tensor. This is an ordinary connection $\nabla$ with torsion on $M$. In a coordinate system, its coefficients are $\G^{i}_{jk}=\mathring{\G}^{i}_{jk}-\frac 12 \Pi^{il}H_{jkl}$, where $\mathring{\G}^{i}_{jk}$ are symmetric in the lower indices (they correspond to the coefficients of a connection $\mathring{\nabla}$ without torsion) and therefore the torsion of $\nabla$ in component reads $\Pi^{il}H_{jkl}\equiv H^{i}{}_{jk}$. (We introduce a notation different than \cite{Ikeda:2019czt} and \cite{Chatzistavrakidis:2021nom}, where this torsion tensor is denoted as $\Theta$, maintaining the logic that indices of differential form twists like $H_3$ are raised with the map $\Pi^{\sharp}$.)  Moreover, note that a dual connection is naturally induced on $T^{\ast}M$, usually denoted as $\nabla^{\ast}$ and having coefficients $-\G^{i}_{jk}$. 

Aside the above ordinary connection, we also have to consider a non-ordinary one, defined as follows. 
Recall that a connection on a vector bundle $V$ is a linear map $\nabla:\G(V)\to \G(T^{\ast}M\otimes V)$ that satisfies the Leibniz rule 
\be 
\nabla(f v)= f\nabla v +\dd f\otimes v\,, \quad v\in\G(V), f\in C^{\infty}(M)\,.
\ee  
Associated to it is a covariant derivative $\nabla_{X}:\G(V)\to \G(V)$ along a vector field $X\in \G(TM)$, which satisfies the usual linearity conditions and the Leibniz rule 
\be \label{Leibnizordinary}
\nabla_{X}(f v)=f\nabla_{X}v+X(f) v\,. 
\ee 
These two common notions may be generalized to the case when $X$ is not a vector field but instead a section of some suitable vector bundle $E$, such as a Lie algebroid. Specifically, an $E$-connection on $V$ is then a linear map $^{E}\nabla: \G(V) \to \G(E^{\ast}\otimes V)$ and as such the associated $E$-covariant derivative $^{E}\nabla_{e}$ with argument $e\in\G(E)$ satisfies a Leibniz rule with the help of the anchor map $\rho$:
\bea 
^{E}\nabla_{e}(f v)=f\, ^{E}\nabla_{e}v +\rho(e)f \, v\,.
\eea 
Note that $V$ can also have a Lie algebroid structure, in particular it can be $E$ itself, in which case one would have an $E$-connection on $E$. To summarize and avoid any potential confusion, one should keep in mind that there are three different notions at play here: (i) an ordinary connection on $V$, (ii) an $E$-connection on $V$ and (iii) the special case of an $E$-connection on $E$. 

Like their ordinary counterparts, $E$-covariant derivatives on $E$ may have a non-vanishing $E$-torsion and $E$-curvature tensors, defined in a straightforward way as 
\bea 
\label{Etorsion} ^{E}T(e,e')&=& \, ^{E}\nabla_{e}e'- \, ^{E}\nabla_{e'}e-[e,e']_E\,, \\[4pt] 
^{E}R(e,e')&=& [^{E}\nabla_{e}, \, ^{E}\nabla_{e'}]- \, ^{E}\nabla_{[e,e']_E}\,,
\eea 
and satisfying the usual properties such as $^{E}T(e,e')=- \, ^{E}T(e',e)$, $^{E}T(f e,e')=fT(e,e')$ etc. 
Furthermore, a useful notion that we will need in this paper is the basic curvature{\footnote{As explained in \cite{Kotov:2016lpx}, this is the curvature of the splitting of the short exact sequence associated to every vector bundle.}} ${\cal S}$, defined as 
\be 
{\cal S}(e,e')={\cal L}_{e}(\nabla e')-{\cal L}_{e'}(\nabla e)-\nabla_{\rho(\nabla e)}e'+\nabla_{\rho(\nabla e')}e-\nabla[e,e']_{E}\,,
\ee  
for the ordinary connection $\nabla$ on $E$. It turns out that this basic curvature is given as \cite{Kotov:2016lpx} 
\be \label{basiccurvature}
{\cal S}=\nabla(^{E}T)+2\text{Alt}\langle\rho,{\cal R}\rangle\,, 
\ee 
where ${\cal R}$ is the ordinary curvature of $\nabla$. We will encounter this formula repeatedly in the field theory setting of twisted R-Poisson sigma models. 

Returning to the discussion of the invariant geometrical meaning of the gauge transformations and field equations, the target space covariant result is 
\bea
\d^{\nabla}A=\DD \epsilon-T(A,\epsilon) \quad \text{and} \quad G^{\nabla}=\DD A-\frac 12 T(A,A)\,, 
\eea  
where $\DD$ is the induced differential by $\nabla$ on the exterior algebra and $T$ is the $E$-torsion of the $E$-connection 
\be 
^{E}\nabla_{e}e':=\nabla_{\Pi(e)}e'
\ee 
for the Lie algebroid $(T^{\ast}M,\Pi^{\sharp},[\cdot,\cdot]_{K})$ encountered before and it is given as
\be 
T=-\mathring{\nabla}\Pi\,. 
\ee 
Note that the $E$-torsion does not contain $H_3$ at all, since it is controlled by the connection \emph{without} torsion $\mathring{\nabla}$. All the $H_3$ dependence is through the differential $\DD$ \cite{Ikeda:2019czt}. 

\subsection{Twisted R-Poisson in any dimension}
\label{sec22}

After this brief introduction to twisted Poisson manifolds, their corresponding Lie algebroids and the associated topological field theory, let us move on to the main concept underlying the field theories we consider in this contribution, namely twisted R-Poisson manifolds. These are equipped with the additional structure of a multivector field $R$ of order $p+1$. This can give rise to a bracket that generalizes the Poisson bracket{\footnote{Note that every multivector field defines a multiderivation on a manifold, namely a multilinear map $C^{\infty}(M)\times\dots C^{\infty}(M)\to C^{\infty}(M)$ which is totally antisymmetric and a $C^{\infty}$-derivation in each of the arguments \cite{Pham}. In addition, the space of multiderivations and the space of multivector fields of the same order are in one-to-one correspondence.}} $\{\cdot,\cdot\}$, however we directly describe the structure in terms of the alternative formulation based on the Schouten-Nijenhuis bracket. Therefore, we consider the quadruple $(M,\Pi,R,H)$ consisting of a smooth manifold $M$ equipped with a bivector $\Pi\in\G(\wedge^{2}TM)$, an antisymmetric multivector $R\in \G(\wedge^{p+1}TM)$ of degree $p+1$  
and a $(p+2)$-form $H\in\G(\wedge^{p+2}T^{\ast}M)$. This is called a twisted R-Poisson manifold when the following conditions hold \cite{Chatzistavrakidis:2021nom}{\footnote{Note a sign correction in the second equation with respect to the original paper.}}
\be \label{twistedRPoisson}
[\Pi,\Pi]_{\text{SN}}=0\,, \quad [\Pi,R]_{\text{SN}}=(-1)^{p+1}\langle \otimes^{p+2}\Pi,H\rangle \quad \text{and} \quad \dd H=0\,.
\ee 
For completeness and in absence of any other structure one may include in the definition the requirement that $[R,R]_{\text{SN}}=0$, although this does not appear in the field theoretic incarnation of twisted R-Poisson target spaces. Evidently, for vanishing $H$ this should be called an (ordinary or untwisted) R-Poisson structure. Moreover, when in addition $R$ is absent, this reduces to a standard Poisson structure. Note that twisted Poisson structures are \emph{not} included in this class; we explain how they appear from this point of view in Section \ref{sec23}.
	
Much like the simpler case of Poisson structures, twisted R-Poisson structures give rise to topological field theory. However, the corresponding topological sigma models are theories defined in $p+1$ dimensions with $p\ge 2$. This can also be seen from the fact that the twist $H$ is a $(p+2)$-form and it appears in the field theory as a Wess-Zumino term. The field content of the topological field theory is chosen such that it can accommodate a 2-vector background and it reads
\be 
(X^{i},A_i,Y^{i},Z_i)\,,
\ee 
with differential form degrees $(0,1,p-1,p)$ respectively and fields taking values in suitable bundles pulled-back via the map $X: \S_{p+1}\to M$ whose components are the real scalars on the $(p+1)$D worldvolume $\S_{p+1}$, namely
\be 
A\in \Omega^{1}(\S_{p+1},X^{\ast}T^{\ast}M) \quad Y\in \Omega^{p-1}(\S_{p+1},X^{\ast}TM) \quad Z\in \Omega^{p}(\S_{p+1},X^{\ast}T^{\ast}M)\,.
\ee 
To write down an action functional, we include all admissible terms for arbitrary $p$. We emphasize that in spacial number of dimensions there might exist more admissible terms, as discussed in Section \ref{sec23}. The generic action functional for $p\ge 2$ is 
	\bea 
	S^{(p+1)}&=& \int_{\S_{p+1}}\left(Z_i\w\dd X^{i}-A_i\w \dd Y^{i}+\Pi^{ij}(X)\, Z_i\w A_j-\frac 12 \, \partial_{k}\Pi^{ij}(X)\, Y^{k}\w A_i\w A_j\, \,+ \right. \nn \\[4pt] && \quad \qquad \left. +\, \frac 1{(p+1)!}R^{i_1\dots i_{p+1}}(X)\, A_{i_1}\w\dots \wedge A_{i_{p+1}}\right)+\int_{\S_{p+2}}X^{\ast}H\,. \label{Sp+1}\nn
	\eea  
The gauge symmetries of the theory are found to be 
\bea 
\label{gt1} \d X^{i}&=&\Pi^{ji}\epsilon_{j}\,,\\[4pt]
\label{gt2} \d A_i&=&\dd\epsilon_i+\partial_i\Pi^{jk}A_j\epsilon_k \,,\\[4pt]
\d Y^{i}&=&(-1)^{p-1}\dd\chi^{i}+\Pi^{ji}\, \psi_j -\partial_j\Pi^{ik}\left(\chi^{j}A_k+Y^{j}\epsilon_k\right)+\frac 1{(p-1)!}R^{iji_1\dots i_{p-1}}A_{i_1}\dots A_{i_{p-1}}\epsilon_j\,, \\[4pt] \label{gt3}
\d Z_i&=&(-1)^{p}\dd\psi_i+\partial_i\Pi^{jk}\left(Z_j\epsilon_k+\psi_jA_k\right) -\partial_i\partial_j\Pi^{kl}\left(Y^jA_k\epsilon_l-\frac 12 \, A_kA_l\chi^{j}\right) \, + \label{gt4}\\[4pt] 
&   +&  \frac {(-1)^p}{p!} \, \partial_iR^{ji_1\dots i_{p}}A_{i_1}\dots A_{i_{p}}\epsilon_j\,{-\,\frac 1{(p+1)!}\Pi^{kj}H_{ijl_1\dots l_p}\epsilon_k\sum_{r=1}^{p+1}(-1)^{r}\prod_{s=1}^{r-1}\dd X^{l_{s}}\prod_{t=r}^{p}\Pi^{l_tm_t}A_{m_t} }\,, \nn
\eea 
in terms of 3 gauge parameters $\e_i,\chi^{i},\psi_i$ of differential form degree $(0,p-2,p-1)$ respectively. Note that the last term in the gauge transformation of the highest-form field $Z_i$ is the only one that contains the components of the twist $H$. It appears together with all admissible combinations of $\dd X$ and $A$ with suitably alternating sign so as to cancel the contribution of the Wess-Zumino term in the gauge variation of the action functional. As such, it is the generalization to higher dimensions of the corresponding term in the gauge transformation of the 1-form in the twisted Poisson sigma model. 
In addition, one can determine the four classical field equations of the theory. We will not write them explicitly here, their precise form may be found in the original paper \cite{Chatzistavrakidis:2021nom}, 
however for notation purposes we write
\be 
\{F^{i}\supset \dd X^{i}\,, \,\, G_{i}\supset\dd A_i\,,\,\, {\cal F}^{i}\supset\dd Y^{i}\,,\,\, {\cal G}_{i} \supset\dd Z_{i}\} \, =  0\,,
\ee 
by which we mean that the four field strengths appearing in the curly brackets start with the differential on the corresponding field, they contain multiple nonlinear terms that we do not present and the field equations are that they vanish. 

The theory introduced above is a highly reducible constrained Hamiltonian system. Since some of the fundamental fields are differential forms of degree higher than 1 and there are gauge parameters that are not scalars, the gauge parameters are not all independent. In addition, the gauge algebra of the theory is open, even in the absence of the twist $H$, which is not surprising. What is somewhat unusual is that in the present case the commutator of gauge transformations does not only contain linear terms in the field equations but also nonlinear ones. To illustrate this ``nonlinear openness'', we note for instance that the commutator of two gauge transformations on $Z_i$ contains on the right hand side a variety of terms, 
\be 
[\d_1,\d_2]Z_i = \d_{12}Z_i + (\dots)_i^{j}\, G_j + (\dots)_{ij}{\cal F}^{j}+(\dots)_{ij}F^{j} +(\dots)_{ijk}{F^{j} F^{k}}+\dots +(\dots)_{ij_1...j_p}{F^{j_1}\dots F^{j_p}}\,,\nn
\ee
where the ellipses denote known combinations of the background quantities, namely the components of $\Pi$ and $R$ and their derivatives. Note also that only the field strength $F$ appears nonlinearly.  

One may now ask what is the target space covariant formulation of this class of theories. Unlike the twisted Poisson sigma model, in the present case we already encounter quantities that are not manifestly tensorial in the action functional. This is once more the partial derivative on the components of the Poisson structure $\Pi$. What is more, the gauge transformations and the field equations contain second derivatives on $\Pi^{ij}$ and also derivatives on the components of $R$. To properly give an invariant meaning to all these, first of all we should introduce again a connection on $TM$, this time without torsion. We denote it again as $\mathring{\nabla}$. This induces a dual connection on $T^{\ast}M$ and also a covariant differential $\mathring{\DD}$ on the exterior algebra of differential forms. 
A well-defined procedure than leads to all the covariant expressions for the gauge transformations and field equations for all fields of the theory. Indicatively, let us present here two illustrative ones, the field equations with respect to $A_i$ and $X^{i}$---the rest are found in the original paper:
	\bea 
	{\cal F}^{\mathring{\nabla}}&=&\mathring{\DD}Y-T(A,Y)+(-1)^{p}\Pi({Z}^{\mathring{\nabla}})-\frac 1{p!}R(A,\dots,A)\,,\\[4pt]
	{\cal G}^{\mathring{\nabla}}&=&(-1)^{p+1}\mathring{\DD}{Z}^{\mathring{\nabla}}-T({Z}^{\mathring{\nabla}},A)+ \frac 12 {\cal S}(Y,A,A)+\frac 1{(p+1)!}(\mathring{\nabla}R+{\cal T})(A,\dots,A)\,,
	\eea 
	where we introduced the field redefinition 	${Z}^{\mathring{\nabla}}_i=Z_i+\mathring{\G}_{ij}^{k}Y^{j}\w A_k$. 
	Remarkably, we observe that tensors of $E$-geometry for $E=T^{\ast}M$ appear again in these covariant expressions. In particular we encounter the $E$-torsion $T=-\mathring{\nabla}\Pi$ in both expressions. Furthermore, we notice the appearance of the basic curvature ${\cal S}$ in ${\cal G}^{\mathring{\nabla}}$, given in Eq. \eqref{basiccurvature}. This is the covariantization of the double partial derivative on the components of the Poisson structure. Note that in the case of the twisted Poisson sigma model, the basic curvature did not appear at all at this level; it appears only when one computes the gauge algebra or, equivalently, the square of the BRST operator. In the twisted R-Poisson case it appears already at the level of the field equations. Finally, note the appearance of the tensors $\mathring{\nabla}R$ and ${\cal T}\in\G(T^{\ast}M\otimes \bigwedge^{p+1}TM) $. The latter is defined as 
	\be 
	{{\cal T}:=\langle\otimes^{p+1}\Pi,H_{p+2}\rangle }
	\ee
	and it is the higher analogue of the torsion of the \emph{ordinary} connection in the twisted Poisson case. A precise geometrical description of this higher torsion tensor will be given elsewhere. 
	
	Finally, the action functional itself can be written in a fully covariant form as follows: 
	\begin{align}
\label{Sp+1cov}
S^{(p+1)}  = \int_{\S_{p+1}}\left(\langle {Z}^{\mathring{\nabla}},F\rangle-\langle Y,G^{\mathring{\nabla}}\rangle+\frac 1{(p+1)!}R(A,\dots,A)\right)+\int_{\S_{p+2}}X^{\ast}H\,.   
\end{align}
This provides the geometric completion of local patch results for string and M-theory fluxes corresponding to Figure \ref{t1} \cite{Mylonas:2012pg,Chatzistavrakidis:2019seu,Chatzistavrakidis:2018ztm,Bessho:2015tkk,Heller:2016abk}. 
	
\subsection{Bi-twisted R-Poisson and other deformations}
\label{sec23}

As mentioned in the previous section, there exist additional special cases with the same field content that are not covered by the above analysis. One can view those as deformation of the twisted R-Poisson sigma models that arise in special dimensions, in particular in 2, 3 and 4 dimensions. As found in \cite{Chatzistavrakidis:2021nom}, these deformations appear in Table \ref{t2}. 

\begin{table}\centering
	\begin{tabular}{ |c||c| }
		\hline
		{$\text{dim}\, \S_{p+1}$} & Admissible deformations  \\ 
		\hline  
		\multirow{3}{0.5em}{2} &\\ & $f^{ij}(X,Y)\,Z_i\w Z_j$, \, $f^{ij}(Y)\,A_i\w A_j$  \\  & \\
		\hline 
		\multirow{3}{0.5em}{3} &\\ & $f_{ijk}(X)Y^{i}\w Y^{j}\w Y^{k}$, \, $f^{i}_{j}(X)\, Z_i\w Y^{j}$,\, $f_{ij}^{k}(X)Y^{i}\w Y^{j}\w A_k$\\ &\\ \hline 
		\multirow{3}{0.5em}{4} &\\ &  $f_{ij}(X)Y^i\w Y^j$ \\& \\ \hline 
	\end{tabular} \caption{Admissible deformations of twisted R-Poisson sigma models}\label{t2}
\end{table} 

Let us briefly summarize the type of theories one obtains through these deformations. In 2 dimensions ($p=1$), generically we find ``doubled'' sigma models, in other words sigma models with two sets of real scalar fields, since one now has that the $(p-1)$-form $Y^{i}$ is a scalar. Therefore, this leads to much more options, out of which one is simply the twisted R-Poisson sigma model already discussed. Interestingly, a different choice of parameters leads to the twisted Poisson sigma model coupled to 2D BF theory.  Thus we see that the twisted Poisson sigma model discussed in Section \ref{sec21}
is also included in our analysis as a deformation of the twisted R-Poisson sigma model in 2D. 

In 3 dimensions we encounter the most interesting case. As already discussed in the introduction, Courant sigma models constitute a very general class of topological field theories in 3 dimensions. They contain Chern-Simons theory as a special case, but also untwisted R-Poisson sigma models. Twisted R-Poisson sigma models would then be a special case of 4-form-twisted Courant sigma models \cite{Hansen:2009zd}. Inspecting the deformations of Table \ref{t2}, we observe that they are precisely the ones that lead to Courant structures, see \cite{Chatzistavrakidis:2021nom} for a more detailed explanation.  In the present context, $R$ is a trivector. The special value 3 for its vector degree matches the one of the Schouten-Nijenhuis bracket of $\Pi$ with itself. This is only possible in 3 dimensions. Hence, one may use $R$ to twist the Poisson substructure of twisted R-Poisson, thus being led in a bi-twisted structure. Specifically, a bi-twisted R-Poisson structure is a quintuple $(M,\Pi,R,S,H)$ of a smooth manifold $M$ equipped with a 2-vector $\Pi$, a 3-vector $R$, a 3-form $S$ and a 4-form $H$ such that 
\bea \label{bitRpoisson}
	\frac 12 \, [\Pi,\Pi]&=&R-\langle \Pi\otimes \Pi\otimes \Pi,S\rangle\,,\\[4pt] 
	\dd S&=&H\,.
	\eea 
Note that for $R=0=H$ the structure reduces to $S$-twisted Poisson. In general, this leads to a bi-twisted R-Poisson sigma model, whose covariant formulation requires a connection with torsion. 

Finally, we mention that in 4 dimensions the deformation is a ``'theta-term'' type of contribution. Such models were considered before in Ref. \cite{Ikeda:2010vz}. 

\section{Differential graded manifold perspective \& BV}
\label{sec3}

A differential graded manifold or simply Q-manifold is a graded supermanifold equipped with a cohomological vector field $Q$, for which 
\be 
Q^2=\frac 12 \{Q,Q\}=0\,.
\ee 
Remarkably, there exists an exact correspondence between Q-manifolds and Lie algebroids \cite{Vaintrob}. Indeed, instead of thinking of a Lie algebroid as the triple $(E,\rho,[\cdot,\cdot]_E)$ of a vector bundle with an anchor map and a binary skew-symmetric operation satisfying the Leibniz rule, one can think of it in a significantly simpler way as a Q-manifold $(E[1],Q_E)$, where $E[1]$ is the degree-1-shifted vector bundle $E$. 

In Section \ref{sec21}, we encountered the twisted Poisson structure $(M,\Pi,H_3)$ and the Lie algebroid $(T^{\ast}M,\Pi^{\sharp},[\cdot,\cdot]_K)$. In the present formulation, one can introduce the graded supermanifold $T^{\ast}[1]M$, with local coordinates $x^{i}$ of degree 0 and $a_i$ of degree 1. Let us denote by $\partial_{x^{i}}$ and $\partial_{a_i}$ the associated derivatives $\partial/\partial x^{i}$ and $\partial/\partial a_i$ respectively. Then consider a vector field of degree 1 on this graded manifold of the form
\be 
Q_{T^{\ast}M}=\Pi^{ij}(x)\xi_i\partial_{x^{j}}-\frac 12 (\partial_i\Pi^{jk}+\Pi^{jl}\Pi^{km}H_{ilm})\xi_{j}\xi_{k}\partial_{\xi_{i}}\,.
\ee 
Then, this vector field is cohomological if and only if the twisted Poisson condition \eqref{twistedPoisson} holds. 

In the case of twisted R-Poisson structure a Q-manifold can be found in a similar way. This time one begins with the graded manifold $T^{\ast}[1]M$ of the previous paragraph and considers its cotangent bundle with shifted degree by $p$, namely $T^{\ast}[p]T^{\ast}[1]M$. As explained already in the introduction, this graded manifold can be assigned coordinates of degrees $(0,1,p-1,p)$, which we denote respectively as $(x^i,a_i,y^i,z_i)$. The notation should already indicate that the fields of the twisted R-Poisson sigma model are the pull-backs of these four sets of graded coordinates by some general sigma model map. We now consider the following degree-1 vector field on the graded manifold: 
\bea  
&& Q=\Pi^{ji}a_j\partial_{x^{i}}-\frac 12 \partial_i\Pi^{jk}a_ja_k\partial_{a_{i}}+\left((-1)^{p}\Pi^{ji}z_j-\partial_j\Pi^{ik}a_ky^{j}+\frac 1{p!}R^{ij_1\dots j_p}a_{j_1}\dots a_{j_p}\right)\partial_{y^{i}} \, + \nn\\[4pt] 
&& \qquad +\, \left(\partial_i\Pi^{jk}a_kz_j-\frac {(-1)^{p}}2 \partial_i\partial_j\Pi^{kl}y^{j}a_ka_l+\frac {(-1)^{p}}{(p+1)!} f_{i}^{k_1\dots k_{p+1}}a_{k_1}\dots a_{k_{p+1}}\right)\partial_{z_{i}}\,,\eea
where we have defined 
\be 
f_{i}^{k_1\dots k_{p+1}}= \partial_iR^{k_1\dots k_{p+1}}+\prod_{r=1}^{p+1}\Pi^{k_rl_r}H_{il_1\dots l_{p+1}}\,.
\ee
Then $(T^{\ast}[p]T^{\ast}[1]M,Q)$ is a Q-manifold if and only if the defining conditions \eqref{twistedRPoisson} for a twisted R-Poisson structure hold. Thus we have described twisted R-Poisson manifolds as Q-manifolds. 

Note now that the Q-manifold in both cases above is a cotangent bundle---a second order cotangent bundle in the latter case. This means that apart from its structure as a Q-manifold (``Q-structure''), it also has a natural graded symplectic structure (``P-structure''). In Darboux coordinates this is simply given by a graded symplectic 2-form, say $\omega$, which is of the general form $\dd(\text{coordinates})\w\dd (\text{momenta})$. In the case of $T^{\ast}[1]$, the coordinates are $x^i$ and the corresponding momenta are $a_i$. The symplectic form in this case is of degree 1. In the case of $T^{\ast}[p]T^{\ast}[1]M$, we can consider the $x^i$ and $a_i$ as coordinates of the second order cotangent bundle and correspondingly $z_i$ and $y^i$ as the associated momenta. The symplectic form is then of degree $p$. 

The above discussion leads to the notion of a QPp-manifold \cite{Schwarz:1992nx}, a graded manifold equipped with a cohomological vector field and a graded symplectic 2-form $\omega$ of degree $p$ such that the latter is invariant under the flow of the former, specifically 
\be \label{compatibility}
{\cal L}_{Q}\omega=0\,.
\ee 
Notable examples of QP-manifolds are the ones corresponding to Poisson geometry and Courant algebroids. More examples may be found e.g. in the review \cite{Ikeda:2012pv}. Note though the absence of the word ``twisted'' in the aforementioned examples. Twisted Poisson and twisted Courant (including twisted R-Poisson) structures do not satisfy the condition \eqref{compatibility} exactly. For twisted R-Poisson structures one finds that the right hand side in the Lie derivative along $Q$ is an exact form, 
\be 
{\cal L}_{Q}\omega|_{\text{R-Poisson}} \propto \dd(\iota_{\Pi^{\sharp}}^{p+1}H)\,,
\ee 
where the notation $\iota^{p+1}$ means the interior product taken $p+1$ times. This is essentially the same as the tensor ${\cal T}$ of Section \ref{sec22}. As long as it is not closed, there exists an obstruction to the QP-ness of the geometrical structure on $T^{\ast}[p]T^{\ast}[1]M$. This means that in general 
\bi 
\item Wess-Zumino terms obstruct QP-ness even in the presence of both Q and P structures.
\ei 

To understand the importance of this statement, let us briefly shift our attention to the BV quantization of this class of topological field theories. As is well-known, quantization in the BV formalism is necessary when we encounter gauge theories that have one or more of the following properties: 
\bi 
\item The gauge algebra closes only when the classical equations of motion are used.
\item The gauge algebra has structure functions that depend on the fields of the theory.
\item The gauge transformations are reducible.
\ei 
The theories we discuss exhibit all three properties. 

In applying the classical BV formalism to a theory with a given action functional \cite{HT,Gomis:1994he}, say $S_0$, and a set of gauge symmetries, first one enlarges the configuration space by ghosts, ghosts for ghosts and antifields. Ghosts are in correspondence with the gauge parameters and ghosts for ghosts take care of the dependencies among the gauge transformations, in other words of the reducibility. For each field, ghosts and ghosts for ghost there is a corresponding antifield. Then one defines an odd symplectic structure on the space of all these fields and antifields, the BV (anti)bracket $(\cdot,\cdot)_{\text{BV}}$. Next the classical action $S_0$ is extended to one with all admissible terms with ghosts and antifields, say $S$. Finally one should solve the classical master equation 
\be 
(S,S)_{\text{BV}}=0\,,
\ee 
subject to suitable boundary conditions. The solution of the classical master equation, the BV action, is unique up to canonical transformations. Note that the same information can be encoded in the BV operator $s$. Recalling that the square of the BRST operator $s_0$ on fields does not vanish, but instead is proportional to the classical field equations, the BV operator is its extension by antifield terms such that it is nilpotent off-shell. It is related to the classical BV action $S$ via 
\be 
s\varphi=(S,\varphi)_{\text{BV}}\,,
\ee 
for any field $\varphi$ in the theory. In fact, it is often computationally easier to determine $s$ rather than $S$. 

Instead of the traditional approach sketched above, there exists a geometrical method to determine the solution to the classical master equation. This is called the AKSZ construction \cite{Alexandrov:1995kv} and it amounts to considering maps from a Q-manifold $\widehat{\Sigma}$ (typically being $T[1]\S$, but not necessarily) to a QP$p$-manifold ${\cal M}$ along with a degree $p+1$ Hamiltonian function $\Theta\in C^{\infty}({\cal M})$ such that 
\be 
Q=\{\Theta,\cdot\}\,,
\ee 
where $Q$ refers to the cohomological vector field on the target ${\cal M}$ and the curly brackets denote the induced Poisson bracket by the symplectic form on ${\cal M}$. Due to the nilpotency of $Q$, the Hamiltonian function satisfies 
\be 
\{\Theta,\Theta\}=0\,.
\ee 
This is a key equation, since it guarantees that once the Hamiltonian is pulled-back via the pull-back of the map from $\S$ to ${\cal M}$, say $\varphi$, to an action functional 
\be 
S[\varphi]=\int\left(\frac 12 \omega_{ab}\,\varphi^{a}\w \dd\varphi^{b} +\varphi^{\ast}(\Theta)\right)\,,
\ee 
then this action functional satisfies the classical master equation. This way we see that geometry and field theory go hand in hand. 

We briefly recall that for the (untwisted) Poisson sigma model in 2D, 
obtained from the discussion in Section \ref{sec21} with $H_3=0$, the target QP1-manifold is $T^{\ast}[1]M$. The full field content appears in Table \ref{t3}. It contains the scalar fields $X^{i}$, the 1-forms $A_i$, the single scalar ghost $\e_i$, which we denote with the same letter as the gauge parameter to avoid clutter, and their corresponding antifields. Note that the antifields have ghost number $\text{gh}(\varphi^+)$ equal to $-1-\text{gh}(\varphi)$ and form degree $f(\varphi^+)=2-f(\varphi)$. 

\begin{table}
	\begin{center}
		\begin{tabular}{|c||c|c|c|c|c|c|}
			\hline
			\multirow{3}{5em}{(Anti)Field} &&&&&& \\ & $X^i$ & $A_i$ & $\e_i$ & $X^{+}_i$ & $A_{+}^{i}$ & $\e_{+} ^{i}$\\ &&&&&& \\
			\hline
			\multirow{3}{3em}{gh($\cdot$)} &&&&&& \\ & 0 & 0 & 1 & -1 & -1 & -2\\ &&&&&&\\
			\hline
		\multirow{3}{3.5em}{Form degree} &&&&&&\\ & 0 & 1 & 0 & 2 & 1 & 2 \\ &&&&&& \\
			\hline
	\end{tabular}\end{center}\caption{Fields and antifields of the Poisson sigma model.}\label{t3}
\end{table}

The worldsheet Q-manifold $T[1]\S_2$ has coordinates $\s^{\a},\theta^{\a}$ of degree 0 and 1 respectively. Then one can define two superfields: 
\bea
\mathbf{X}^i&=&X^i+A^{ i}_{+\a}\theta^{\a}-\sfrac 12 \e^{ i}_{+\a\b}\th^{\a}\th^{\b}~, \nn\\ 
\mathbf{A}_i&=&\e_i+A_{i\a}\th^{\a}+\sfrac 12 X^{+}_{i\a\b}\th^{\a}\th^{\b}~.\nn
\eea 
The
BV action is given in terms of them as 
\be 
S=\int \left(\mathbf{A}_i\, \dd \mathbf{X}^{i}+\frac 12 \, \Pi^{ij}(\mathbf{X})\,\mathbf{A}_i \mathbf{A}_j\right)\,, \nn
\ee  	 
which is simply the ``bold'' version of the classical action. 
A similar approach works in the 3D case of general Courant sigma models, including (untwisted) R-Poisson ones. Once more, the action is the ``bold'' version of the corresponding classical one, once the correct superfields are defined for the four fields $X^{i}, A_i, Y^{i}, Z_i$. 

Once Wess-Zumino terms are turned on, the above logic does not apply any longer.  This was demonstrated explicitly in \cite{Ikeda:2019czt}, where the authors showed that the naive generalization of the AKSZ action does not work for the 3-form-twisted Poisson sigma model in 2D. In other words, the ``bold'' action is not the correct one. The underlying reason is the obstruction to QP-ness that we discussed earlier. Nevertheless, the correct BV action can be identified in a traditional way, as described in the beginning. This is what \cite{Ikeda:2019czt} did. 

On the other hand, the 3-form-twisted Poisson sigma model in 2D is only a single example. How special is it? The construction of twisted R-Poisson sigma models was essentially motivated by this question. Note also that the latter theories exhibits features that do not appear in the 2D case. These include the presence of higher differential form gauge parameters (only a scalar gauge parameter appears in 2D) and the nonlinear openness of the gauge algebra (which is linear in 2D).  

Since twisted R-Poisson structures also have an obstruction to the QP structure on ${\cal M}$, one should follow the traditional BV formalism, which in the present case is more demanding. The complete field content of the theory appears in Tables \ref{t4} and \ref{t5}. There we see that there are two towers of ghosts for ghosts, and the corresponding antifields. For the $\chi$-series of ghosts for ghosts the number $r$ runs from $0$ to $p-2$, whereas for the $\psi$-series from $0$ to $p-1$. The ghosts for ghosts with the highest ghost degree are scalars. 

\begin{table}
	\begin{center}	\begin{tabular}{| c | c | c | c | c | c | c | c |}
			\hline 
			\multirow{3}{5.2em}{Field/Ghost} &&&&&&& \\ & $X^{i}$ & $A_i$ & $Y^i$ & $Z_{i}$ & $\epsilon_i$ & $\chi^{i}_{(r)}$ & $\psi_i^{(r)}$ \\ &&&&&&& \\ \hhline{|=|=|=|=|=|=|=|=|}
			\multirow{3}{6.0em}{Ghost degree} &&&&&&& \\ & $0$ & $0$ & $0$ & $0$ & $1$ & $r+1$ & $r+1$ \\ &&&&&&& \\\hline 
			\multirow{3}{5.5em}{Form degree} &&&&&&& \\  & $0$ & $1$ & $p-1$ & $p$ & $0$ & $p-2-r$ & $p-1-r$ \\ &&&&&&&
			\\\hline 
	\end{tabular}\caption{The fields of the twisted R-Poisson sigma model in $p+1$ dimensions}\label{t4}\end{center}\end{table}
	
	\begin{table}
	\begin{center}	\begin{tabular}{| c | c | c | c | c | c | c | c |}
			\hline 
			\multirow{3}{4em}{Antifield} &&&&&&& \\ & $X^{+}_{i}$ & $A_{+}^i$ & $Y^{+}_i$ & $Z_{+}^{i}$ & $\epsilon_{+}^i$ & $\chi^{+}_{i}{}^{(r)}$ & $\psi_{+}^i{}_{(r)}$ \\ &&&&&&& \\ \hhline{|=|=|=|=|=|=|=|=|}
			\multirow{3}{6.0em}{Ghost degree} &&&&&&& \\ & $-1$ & $-1$ & $-1$ & $-1$ & $-2$ & $-r-2$ & $-r-2$ \\ &&&&&&& \\\hline 
			\multirow{3}{5.5em}{Form degree} &&&&&&& \\  & $p+1$ & $p$ & $2$ & $1$ & $p+1$ & $r+3$ & $r+2$ \\ &&&&&&&
			\\\hline 
	\end{tabular}\caption{The antifields of the twisted R-Poisson sigma model in $p+1$ dimensions}\label{t5}\end{center}\end{table}

To determine the solution to the classical master equation, one would then consider the most general form of the BV action as an expansion on antifields, 
\be 
S=S^{(0)}+S^{(1)}+\dots +S^{(p+1)}\,,
\ee 
where $S^{(n)}$ contains $n$ antifields. Contrary to the Poisson case in 2D, where $n$ goes up to 2, and to the Courant case in 3D, where $n$ goes up to 3, the general case is difficult. Nevertheless, since the known gauge transformation on the fields produces the BRST operator on them, one can use ``refinements'' by antifields to determine the BV operator on all fields, ghosts and antifields. One can then compare the BV operator to the one found via the AKSZ contruction in the untwisted case and write schematically 
\be 
s\varphi = s_{\tiny{\text{AKSZ}}}\varphi + (\D s\varphi)(H,F)\,,
\ee 
where the discrepant piece is a both $H$-dependent and $F$-dependent, $H$ being the $(p+2)$-form twist and $F$ the field equation of $Z_i$. 
One interesting outcome is that all ghosts in the $\psi$-series, save the highest, receive equation of motion corrections. This strategy was developed recently in \cite{CIS} and it led to complete closed expressions for the BV operator and the BV action  for twisted R-Poisson-Courant sigma models in 3D and also for untwisted ones in any dimension. We refer to the original paper for a detailed discussion.

\section{Conclusions and outlook} 
\label{sec4} 

We close this contribution with a few messages and comments. From the study of twisted R-Poisson sigma models and their underlying geometry, we draw the conclusion that Poisson and twisted Poisson sigma models are not a strictly two-dimensional story. There exist topological field theories in higher dimensions that are based on these geometric structures and extensions thereof by higher degree multivector fields. This includes the special case of bi-twisted R-Poisson structures in 3D, which combine twisted Poisson and twisted R-Poisson in a precise way. It would be interesting to explore how special these bi-twisted structures really are or whether they can somehow appear in higher than 3 dimensions too.   

A second conclusion is that the study of target space covariance for such topological field theories reveals strong ties to the geometry of higher algebroids and in particular of the notions of $E$-connections, $E$-curvature and $E$-torsion for a Lie algebroid vector bundle $E$. Nevertheless, the complete description of the $E$-geometric structure for twisted R-Poisson sigma models is still lacking and we hope to report on this in future work. 

The third conclusion is that Wess-Zumino terms complicate the identification of the classical BV action for topological field theories, since they do not allow direct application of the powerful AKSZ construction. This is due to an obstruction in the compatibility of Q and P structures. Twisted R-Poisson sigma models offer a class of possible examples in general worldvolume dimensions beyond the single example of twisted Poisson in 2D that was studied in \cite{Ikeda:2019czt}. The 3D case was fully solved in Ref. \cite{CIS}.

Aside the obstruction to QP structures mentioned above, there exists a more direct way that a vanilla QP structure can be absent. This happens when there is no P structure to start with. Examples of such situations are Dirac sigma models in 2D \cite{Kotov:2004wz}, which are based on the graded target manifold $E[1]$, with $E$ a Dirac structure of the standard $H$-twisted Courant algebroid. The BV action for such models was found very recently \cite{Chatzistavrakidis:2022wdd}. Furthermore, going slightly beyond Poisson structures, one can consider twisted Jacobi manifolds and their topological field theory in 2D, called twisted Jacobi sigma model \cite{Chatzistavrakidis:2020gpv}---see also \cite{Bascone:2020drt} for the untwisted case.   

Some further interesting directions would be the following. First, although the BV action for a variety of theories that do not have a QP structure on the target is identified, a systematic procedure that refines the AKSZ construction is not yet found. The fact that several examples are now worked out is a good motivation to attempt such a general construction. Aside this, it would be interesting to study also the quantum BV action, especially with regard to deformation quantization. Recall that Poisson sigma models are important also from this perspective. They provide a physical realization of the Kontsevich formality theorem \cite{Kontsevich:1997vb} for deformation quantization of Poisson manifolds. Indeed, it turns out that the star product obtained through the diagrammatic approach of Kontsevich corresponds to a 2-point function in the quantization of the Poisson sigma model on a disk \cite{Cattaneo:1999fm}. It would be interesting to investigate whether twisted R-Poisson sigma models lead to deformation quantization in this sense too. Furthermore, one could try going even beyond twisted R-Poisson structures toward the general framework of homotopy Poisson or P$_{\infty}$ structures \cite{Voronov} on the target space. Finally, it is not only the target space whose diverse structures lead to interesting models, but the worldvolume as well. In most instances in the literature, this is taken to be $T[1]\S$, the degree shifted tangent bundle of an ordinary worldvolume. This only allows for differential forms as fields. An exception is Ref. \cite{Chatzistavrakidis:2019len}, where more general graded spacetime manifolds are considered and gauge theories with mixed symmetry tensor fields are described.

\paragraph{Acknowledgements.} I am grateful to Noriaki Ikeda and Grgur \v{S}imuni\'c for enlightening discussions and collaboration in \cite{CIS}. I would also like to thank Larisa Jonke for suggestions on the manuscript and Peter Schupp and Richard Szabo for discussions on the higher geometry associated to twisted R-Poisson manifolds.  
This work is supported by the Croatian Science Foundation Project ``New Geometries for Gravity and
Spacetime" (IP-2018-01-7615).


\begin{thebibliography}{99}
\bibitem{Chatzistavrakidis:2021nom}
A.~Chatzistavrakidis,
``Topological field theories induced by twisted R-Poisson structure in any dimension,''
JHEP \textbf{09} (2021), 045
doi:10.1007/JHEP09(2021)045
[arXiv:2106.01067 [hep-th]].

\bibitem{SchallerStrobl}
P.~Schaller and T.~Strobl,
``Poisson structure induced (topological) field theories,''
Mod.\ Phys.\ Lett.\ A {\bf 9} (1994) 3129
doi:10.1142/S0217732394002951
[hep-th/9405110].

\bibitem{Ikeda}
N.~Ikeda,
``Two-dimensional gravity and nonlinear gauge theory,''
Annals Phys.\  {\bf 235} (1994) 435
doi:10.1006/aphy.1994.1104
[hep-th/9312059].

\bibitem{Klimcik:2001vg}
C.~Klimcik and T.~Strobl,
``WZW - Poisson manifolds,''
J. Geom. Phys. \textbf{43} (2002), 341-344
doi:10.1016/S0393-0440(02)00027-X
[arXiv:math/0104189 [math.SG]].

\bibitem{Ikeda:2021rir}
N.~Ikeda,
``Higher Dimensional Lie Algebroid Sigma Model with WZ Term,''
Universe \textbf{7} (2021) no.10, 391
doi:10.3390/universe7100391
[arXiv:2109.02858 [hep-th]].

\bibitem{Ikeda:2019czt}
N.~Ikeda and T.~Strobl,
``BV and BFV for the H-twisted Poisson sigma model,''
Annales Henri Poincare \textbf{22} (2021) no.4, 1267-1316
doi:10.1007/s00023-020-00988-0
[arXiv:1912.13511 [hep-th]].

\bibitem{Alexandrov:1995kv}
M.~Alexandrov, A.~Schwarz, O.~Zaboronsky and M.~Kontsevich,
``The Geometry of the master equation and topological quantum field theory,''
Int. J. Mod. Phys. A \textbf{12} (1997), 1405-1429
doi:10.1142/S0217751X97001031
[arXiv:hep-th/9502010 [hep-th]].

\bibitem{Vaintrob}
A. Yu. Vaintrob, 
``Lie algebroids and homological vector fields,''
Russ. Math. Surv. 52 428 (1997)

\bibitem{Ikeda:2000yq}
N.~Ikeda,
``A Deformation of three-dimensional BF theory,''
JHEP \textbf{11} (2000), 009
doi:10.1088/1126-6708/2000/11/009
[arXiv:hep-th/0010096 [hep-th]].

\bibitem{Ikeda:2002wh}
N.~Ikeda,
``Chern-Simons gauge theory coupled with BF theory,''
Int. J. Mod. Phys. A \textbf{18} (2003), 2689-2702
doi:10.1142/S0217751X03015155
 [arXiv:hep-th/0203043 [hep-th]].

\bibitem{Hofman:2002jz}
C.~Hofman and J.~S.~Park,
``BV quantization of topological open membranes,''
Commun. Math. Phys. \textbf{249} (2004), 249-271
doi:10.1007/s00220-004-1106-7
 [arXiv:hep-th/0209214 [hep-th]].

\bibitem{Roytenberg:2006qz}
D.~Roytenberg,
``AKSZ-BV Formalism and Courant Algebroid-induced Topological Field Theories,''
Lett. Math. Phys. \textbf{79} (2007), 143-159
doi:10.1007/s11005-006-0134-y
[arXiv:hep-th/0608150 [hep-th]].

\bibitem{Hansen:2009zd}
M.~Hansen and T.~Strobl,
``First Class Constrained Systems and Twisting of Courant Algebroids by a Closed 4-form,''
Fundamental Interactions, pp. 115-144 (2009)
doi:10.1142/9789814277839\_0008
[arXiv:0904.0711 [hep-th]].

\bibitem{CIS} 
A.~Chatzistavrakidis, N.~Ikeda and G.~\v{S}imuni\'c,
``The BV action of 3D twisted R-Poisson sigma models,''
[arXiv:2206.03683 [hep-th]].



\bibitem{Chatzistavrakidis:2018ztm}
A.~Chatzistavrakidis, L.~Jonke, F.~S.~Khoo and R.~J.~Szabo,
``Double Field Theory and Membrane Sigma-Models,''
JHEP \textbf{07} (2018), 015
doi:10.1007/JHEP07(2018)015
[arXiv:1802.07003 [hep-th]].

\bibitem{Mylonas:2012pg}
D.~Mylonas, P.~Schupp and R.~J.~Szabo,
``Membrane Sigma-Models and Quantization of Non-Geometric Flux Backgrounds,''
JHEP \textbf{09} (2012), 012
doi:10.1007/JHEP09(2012)012
[arXiv:1207.0926 [hep-th]].

\bibitem{Bessho:2015tkk}
T.~Bessho, M.~A.~Heller, N.~Ikeda and S.~Watamura,
``Topological Membranes, Current Algebras and H-flux - R-flux Duality based on Courant Algebroids,''
JHEP \textbf{04} (2016), 170
doi:10.1007/JHEP04(2016)170
[arXiv:1511.03425 [hep-th]].

\bibitem{Heller:2016abk}
M.~A.~Heller, N.~Ikeda and S.~Watamura,
``Unified picture of non-geometric fluxes and T-duality in double field theory via graded symplectic manifolds,''
JHEP \textbf{02} (2017), 078
doi:10.1007/JHEP02(2017)078
[arXiv:1611.08346 [hep-th]].

\bibitem{Chatzistavrakidis:2019seu}
A.~Chatzistavrakidis, L.~Jonke, D.~L\"ust and R.~J.~Szabo,
``Fluxes in Exceptional Field Theory and Threebrane Sigma-Models,''
JHEP \textbf{05} (2019), 055
doi:10.1007/JHEP05(2019)055
[arXiv:1901.07775 [hep-th]].

	\bibitem{Severa:2001qm}
	P.~\v{S}evera and A.~Weinstein,
	``Poisson geometry with a 3 form background,''
	Prog. Theor. Phys. Suppl. \textbf{144} (2001), 145-154
	doi:10.1143/PTPS.144.145
	[arXiv:math/0107133 [math.SG]].
	
\bibitem{Figueroa-OFarrill:2005vws}
J.~M.~Figueroa-O'Farrill and N.~Mohammedi,
``Gauging the Wess-Zumino term of a sigma model with boundary,''
JHEP \textbf{08} (2005), 086
doi:10.1088/1126-6708/2005/08/086
[arXiv:hep-th/0506049 [hep-th]].

	\bibitem{Fulp:2002fm}
	R.~Fulp, T.~Lada and J.~Stasheff,
	``Noether's variational theorem II and the BV formalism,''
	Rend. Circ. Mat. Palermo S \textbf{71} (2003), 115-126
	 [arXiv:math/0204079 [math.QA]].
	
\bibitem{Kotov:2016lpx}
A.~Kotov and T.~Strobl,
``Lie algebroids, gauge theories, and compatible geometrical structures,''
Rev. Math. Phys. \textbf{31} (2018) no.04, 1950015
doi:10.1142/S0129055X19500156
[arXiv:1603.04490 [math.DG]].

\bibitem{Pham}
Pham, D.N., ``Higher Affine Connections,'' Mediterr. J. Math. 13, 1227–1262 (2016). https://doi.org/10.1007/s00009-015-0559-6

 \bibitem{Ikeda:2010vz}
 N.~Ikeda and K.~Uchino,
 ``QP-Structures of Degree 3 and 4D Topological Field Theory,''
 Commun. Math. Phys. \textbf{303} (2011), 317-330
 doi:10.1007/s00220-011-1194-0
 [arXiv:1004.0601 [hep-th]].
 
\bibitem{Schwarz:1992nx}
A.~S.~Schwarz,
``Geometry of Batalin-Vilkovisky quantization,''
Commun. Math. Phys. \textbf{155} (1993), 249-260
doi:10.1007/BF02097392
[arXiv:hep-th/9205088 [hep-th]].

\bibitem{Ikeda:2012pv}
N.~Ikeda,
``Lectures on AKSZ Sigma Models for Physicists,''
doi:10.1142/9789813144613\_0003
[arXiv:1204.3714 [hep-th]].

	\bibitem{HT} 
M. Henneaux and C. Teitelboim,``Quantization of Gauge Systems'', Princeton University Press (1992).

\bibitem{Gomis:1994he}
J.~Gomis, J.~Paris and S.~Samuel,
``Antibracket, antifields and gauge theory quantization,''
Phys. Rept. \textbf{259} (1995), 1-145
doi:10.1016/0370-1573(94)00112-G
[arXiv:hep-th/9412228 [hep-th]].

\bibitem{Kotov:2004wz}
A.~Kotov, P.~Schaller and T.~Strobl,
``Dirac sigma models,''
Commun. Math. Phys. \textbf{260} (2005), 455-480
doi:10.1007/s00220-005-1416-4
[arXiv:hep-th/0411112 [hep-th]].

\bibitem{Chatzistavrakidis:2022wdd}
A.~Chatzistavrakidis, L.~Jonke, T.~Strobl and G.~\v{S}imuni\'c,
``Topological Dirac Sigma Models and the Classical Master Equation,''
[arXiv:2206.14258 [hep-th]].

\bibitem{Chatzistavrakidis:2020gpv}
A.~Chatzistavrakidis and G.~\v{S}imuni\'c,
``Gauged sigma-models with nonclosed 3-form and twisted Jacobi structures,''
JHEP \textbf{11} (2020), 173
doi:10.1007/JHEP11(2020)173
[arXiv:2007.08951 [hep-th]].

\bibitem{Bascone:2020drt}
F.~Bascone, F.~Pezzella and P.~Vitale,
``Jacobi sigma models,''
JHEP \textbf{03} (2021), 110
doi:10.1007/JHEP03(2021)110
[arXiv:2007.12543 [hep-th]].


\bibitem{Kontsevich:1997vb}
M.~Kontsevich,
``Deformation quantization of Poisson manifolds. 1.,''
Lett. Math. Phys. \textbf{66} (2003), 157-216
doi:10.1023/B:MATH.0000027508.00421.bf
[arXiv:q-alg/9709040 [math.QA]].

\bibitem{Cattaneo:1999fm}
A.~S.~Cattaneo and G.~Felder,
``A Path integral approach to the Kontsevich quantization formula,''
Commun. Math. Phys. \textbf{212} (2000), 591-611
doi:10.1007/s002200000229
[arXiv:math/9902090 [math]].

 \bibitem{Voronov} 
 Th. Voronov, 
 ``L-infinity bialgebroids and homotopy Poisson structures on supermanifolds,'' 
 arXiv:1909.04914 [math.DG].
 
 
\bibitem{Chatzistavrakidis:2019len}
A.~Chatzistavrakidis, G.~Karagiannis and P.~Schupp,
``A unified approach to standard and exotic dualizations through graded geometry,''
Commun. Math. Phys. \textbf{378} (2020) no.2, 1157-1201
doi:10.1007/s00220-020-03728-x
[arXiv:1908.11663 [hep-th]].

\end{thebibliography}
\end{document}